\documentclass[11pt]{article}
\usepackage{epsfig}

\begin{document}

\title{The $\Delta$N--Interaction in Hadronic Reactions \\ 
       on a Deuteron Target}
\author{C. A. Mosbacher and F. Osterfeld$^{\dagger}$ \\
Institut f\"ur Kernphysik, Forschungszentrum J\"ulich, \\ 
D--52425 J\"ulich, Germany}
\date{Meson '98 Workshop, Krakow, 30 May -- 2 June 1998} 

\maketitle

\begin{abstract}
The $\pi^+ d$ scattering and the $d(p,n)$ charge exchange 
reaction are used to study the $\Delta N$ interaction in the 
$\Delta$ resonance energy region. Various observables, such 
as inclusive and exclusive 
cross sections, angular distributions, and analyzing powers,
are calculated in a coupled channel approach and shown to
be influenced by the $\Delta N$ potential.
\end{abstract}

In recent years, the $\Delta$ excitation in hadronic reactions
on a deuteron target
has been studied experimentally as well as theoretically.
For the $\pi^+ d$ scattering, the final reaction channels 
$\pi NN$, $\pi d$ and $2p$ have been 
considered within various models 
\cite{maxwell80,garcilazo}.
The coupled channel approach to be introduced here  
can be equally well applied to investigate the $d(p,n)$ 
charge exchange reaction \cite{mosbacher97}.
While the pion excites the $\Delta$ only with spin--longitudinal
coupling, the virtual $\pi$ and $\rho$ meson fields 
produced by the $(p,n)$ system probe both the spin--longitudinal
and the spin--transversal response function. Furthermore,
they obey the energy--momentum relation $\omega^2 < q^2$
and thus explore the $\Delta$ excitation in a region that is
inaccessible to real pion scattering (where $\omega^2 = q^2 + m_\pi^2$). 

\begin{figure}[!t]
\begin{center}
  \epsfig{file=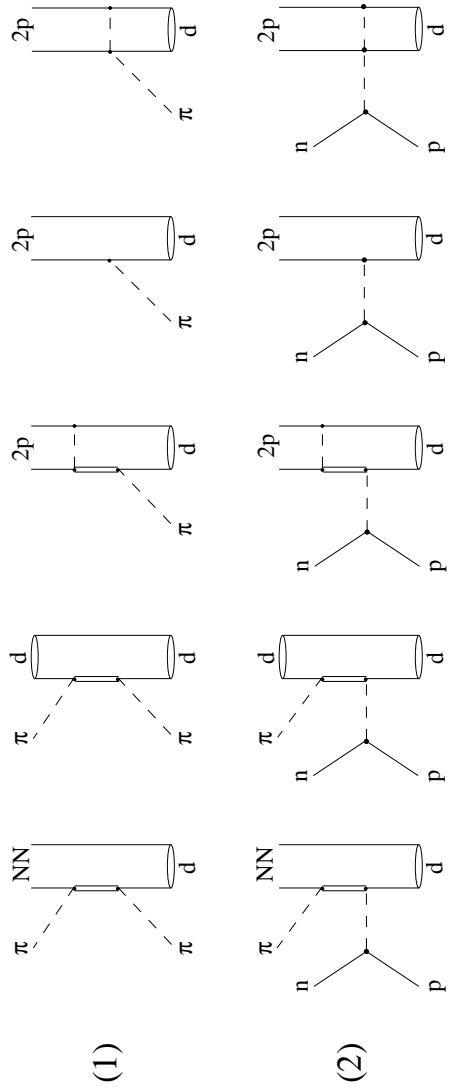, width=6cm}
\end{center}
\caption{Scattering mechanisms for the $\pi^+ d$ reaction (1) 
  and the $d(p,n)$ reaction (2).
  Only the lowest order diagrams are shown, the intermediate 
  $\Delta N$ interaction is not depicted explicitly.} 
\label{fig1}
\end{figure}

The scattering mechanisms considered in our model are represented 
by the diagrams of Fig.~\ref{fig1}. 
The corresponding matrix elements are 
calculated using the source function formalism \cite{udagawa94}.
We set up a system of coupled equations for the correlated 
$\Delta N$ wave function, which includes the effects of $\Delta N$
interactions, and solve the system with the Lanczos method. 
The $\Delta N$ potential 
is constructed in a meson exchange model \cite{machleidt87}
where $\pi$, $\rho$, $\omega$ and $\sigma$ exchange are taken 
into account.
The $\Delta$ resonance is treated thereby as a quasi--particle 
with a given mass and an intrinsic, energy--dependent width. 
After the decay of the $\Delta$, the system turns
into one of the possible final states $\pi NN$, $\pi d$ or $2p$.
For a more detailed explanation of the theoretical framework, please
see Ref.~\cite{mosbacher97}. 

\begin{figure}[!p]
\begin{center}
  \epsfig{file=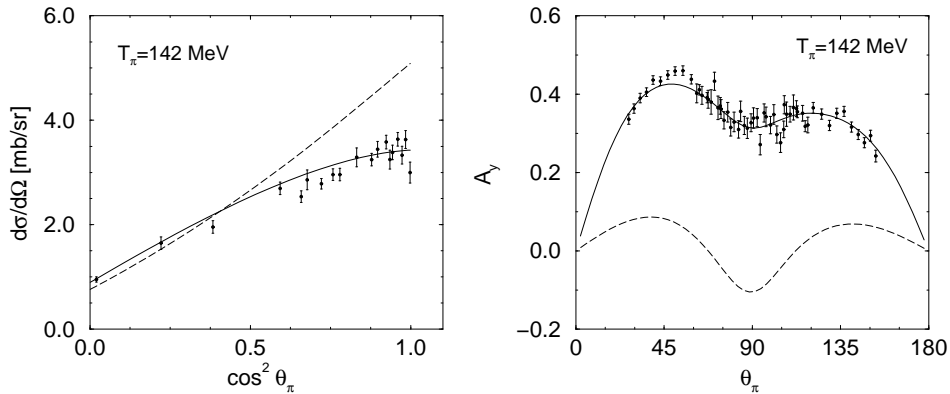, width=12.5cm}
\end{center}
\caption{$d\sigma / d\Omega$ and $A_y$ for the $\pi^+ d \to 2p$
  reaction at $T_\pi = 142$ MeV. Results 
  with (solid) and without (dashed) $V_{\Delta N}$.
  Experimental data are from Refs.~\protect\cite{aebischer76,aprile82}.}
\label{fig2}
\end{figure}

Many observables are influenced by the dynamical treatment of 
the $\Delta N$ system. Exemplary, we present in Fig.~\ref{fig2} 
the differential cross section $d\sigma / d\Omega$ 
and the analyzing power $A_y$
for the pion absorption $\pi^+ d \to 2p$ at $T_\pi = 142$ MeV. 
Solid lines are the result of the full model calculation which 
includes $V_{\Delta N}$ while we put $V_{\Delta N}=0$ to obtain 
the dashed lines. 
Both quantities $d\sigma / d\Omega$ and $A_y$ measure the relative 
strength of the different $\pi d$ partial waves and are therefore 
very sensitive to the $\Delta N$ interaction. 
In the total $\pi d$ cross section 
the effects of the attractive potential become manifest as
a lowering of the $\Delta$ excitation energy, i.e.\ 
as a downward shift of the $\Delta$ resonance peak position.

\begin{figure}[!p]
\begin{center}
  \epsfig{file=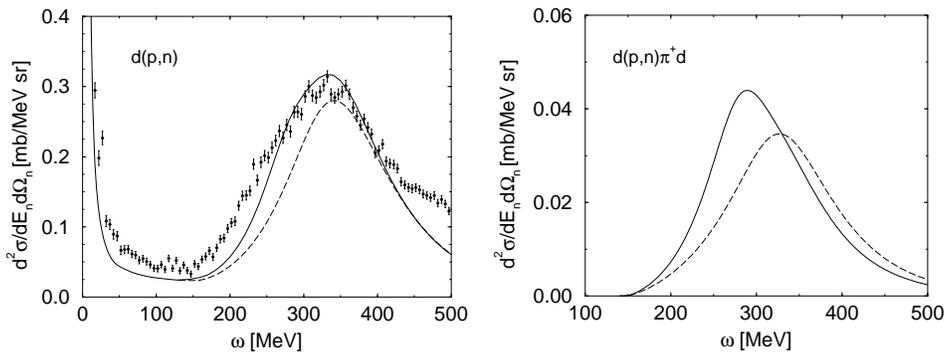, width=12.5cm}
\end{center}
\caption{Inclusive $d(p,n)$ and exclusive $d(p,n)\pi^+ d$ cross section
   at $T_p = 789$ MeV and $\theta_n = 0^o$.
   Results with (solid) and without (dashed) $V_{\Delta N}$. 
   Experimental data are from Ref.~\protect\cite{prout}.}
\label{fig3}
\end{figure}

Fig.~\ref{fig3} shows the total cross section for the inclusive 
$d(p,n)$ reaction and for the exclusive $d(p,n)\pi^ +d$ coherent pion 
production at $T_p=789$ MeV and $\theta_n = 0^o$,
plotted as a function of the energy transfer $\omega = E_p - E_n$.  
The missing inclusive cross section in the dip--region 
(50 MeV $< \omega <$ 250 MeV) is probably due to meson exchange
currents and projectile excitation.
Of special interest is the coherent pion production where both
the peak energy and the magnitude of the cross section 
depend substantially on the strength of $V_{\Delta N}$. 
This is explained by the fact that the $\pi d$ final channel is
strongly coupled to $\Delta N$ states of unnatural parity which are 
subjected to the attractive spin--longitudinal part in the potential.  
Therefore, the exclusive $d(p,n)\pi^+ d$ reaction allows to 
directly measure this attraction and may serve as a tool for further
investigation of the $\Delta N$ interaction.

\end{document}